\documentclass[fleqn,10pt]{wlscirep}

\usepackage{hyperref}
\usepackage{bm}
\usepackage{epsfig}
\usepackage{amssymb}
\usepackage{graphicx}
\usepackage{braket}
\usepackage{bbm}
\usepackage{amsmath}
\usepackage{epsfig}
\usepackage{epstopdf} %EPS files required by SciRep cannot be included using pdfLaTeX without this package
\usepackage{bbold}
\usepackage{dsfont}

\title{Experimental demonstration of a fully inseparable quantum state with nonlocalizable entanglement}

\author[1]{M. Mi\v{c}uda}

\author[1]{D. Koutn\'{y}}

\author[1]{M. Mikov\'{a}}

\author[1]{I. Straka}

\author[1]{M. Je\v{z}ek}

\author[1,*]{L. Mi\v{s}ta, Jr.}

\affil[1]{Department of Optics, Palack\' y University, 17. listopadu 1192/12,  771~46 Olomouc,  Czech Republic}

\affil[*]{mista@optics.upol.cz}

\begin{abstract}
Localizability of entanglement in fully inseparable states is a key ingredient of assisted quantum information protocols as well
as measurement-based models of quantum computing. We investigate the existence of fully inseparable states with nonlocalizable
entanglement, that is, with entanglement which cannot be localized between any pair of subsystems by any measurement on the
remaining part of the system. It is shown, that the nonlocalizable entanglement occurs already in suitable mixtures of a
three-qubit GHZ state and white noise. Further, we generalize this set of states to a two-parametric family of fully inseparable
three-qubit states with nonlocalizable entanglement. Finally, we demonstrate experimentally the existence of nonlocalizable
entanglement by preparing and characterizing one state from the family using correlated single photons and linear optical circuit.
\end{abstract}

\begin{document}

\flushbottom

\maketitle

\thispagestyle{empty}

\section*{Introduction}\label{Sec_1}

Relations among quantum systems can be much more intimate than among everyday classical systems which
we observe through our senses. If two states of two quantum systems with well defined and distinguishable
local properties are superimposed, the individual properties are smeared out whereas the state as a whole
still exhibits well defined global properties. The participating systems are then in a special relationship
because although their local properties are uncertain, they are strongly correlated at same time, whence the
respective states got a fitting nickname entangled states \cite{Schrodinger_35}.

Previous considerations are related only to pure states which
represent an idealization of real states. For more realistic mixed
states the entanglement is defined via a generalization of an
equivalent characterization of pure-state entanglement, that is a
property which cannot be created from pure product states using
only local unitary operations. Thus, generalizing the notion of
local unitaries to the set of local operations and classical
communication (LOCC), which are more sensible operations in the
context of mixed states, we can define a generally mixed entangled
state as a state which cannot be created by LOCC \cite{Werner_89}.
Therefore, from the point of view of entanglement, the set of all
states $\rho_{AB}$ of two subsystems, denoted as $A$ and $B$,
divides into two disjoint subsets. One subset is given by
entangled states whereas the other subset comprise the so called
separable states, i.e., states which can be prepared by LOCC and
therefore attain the following form:
%%%%%%%%%%%%%%%%%%%%%%%%%%%%%%%%%%%%%%%%%%%%%%%%%%%%%%%%%%%%%%%%%%%%%
\begin{equation}\label{separable}
\rho_{AB}=\sum_{i}p_{i}\rho_{A}^{(i)}\otimes\rho_{B}^{(i)},\quad
1\geq p_{i}\geq0,\quad\sum_{i}p_{i}=1,
\end{equation}
%%%%%%%%%%%%%%%%%%%%%%%%%%%%%%%%%%%%%%%%%%%%%%%%%%%%%%%%%%%%%%%%%%%%%%%
where $\rho_{A}^{(i)}$ and $\rho_{B}^{(i)}$ are local states of subsystems $A$ and $B$, respectively.

The discussed entanglement of two quantum systems is a well
understood concept, at least as far as the most simple case of two
systems with two-dimensional state spaces (qubits) is  concerned.
However, the situation changes dramatically for three qubits. This
is because three qubits can be grouped into two groups in three
different ways and therefore we can divide their states into five
classes according to their separability properties with respect to
the particular bipartite splittings \cite{Dur_99}. Thus apart from
fully separable states which can be prepared by LOCC and fully
inseparable states which are entangled across all three
splittings, also partially entangled states exist which have some
splittings separable. Although new phenomena
\cite{Bennett_99,Dur_00,Acin_04} and protocols \cite{Cubitt_03}
can be found also for partially entangled states, most of the
applications utilize fully inseparable states. This involves, for
example, teleportation-based construction of quantum gates
\cite{Gottesman_99} and protocol for quantum secret sharing
\cite{Hillery_99} as well as assisted teleportation
\cite{Karlsson_98} or assisted dense coding \cite{Hao_01}.

All previous applications rely on fully inseparable three-qubit
Greenberger-Horne-Zeilinger (GHZ) state \cite{Greenberger_89}
%%%%%%%%%%%%%%%%%%%%%%%%%%%%%%%%%%%%%%%%%%%%%%%%%%%%%%%%%%%%%%%%%%%%
\begin{equation}\label{GHZ}
|GHZ\rangle_{ABC}=\frac{1}{\sqrt{2}}(|000\rangle_{ABC}+|111\rangle_{ABC}),
\end{equation}
%%%%%%%%%%%%%%%%%%%%%%%%%%%%%%%%%%%%%%%%%%%%%%%%%%%%%%%%%%%%%%%%%%%%
where the first, second and third qubit has been denoted as $A,B$
and $C$, respectively. Except for gate construction
\cite{Gottesman_99} all the applications are based on a genuine
multipartite property of the GHZ state called localizability of
entanglement \cite{Verstraete_04,Popp_05}. Let us imagine, that
three parties called Alice, Bob and Clare hold qubits $A,B$ and
$C$ of GHZ state (\ref{GHZ}), respectively. Now, if Clare performs
a suitable measurement on her qubit $C$, which does not reveal her
any information about which of the two alternatives in the
superposition on the right-hand side (RHS) of Eq.~(\ref{GHZ}) took
place, a maximally entangled state is localized between qubits $A$
and $B$. More precisely, if Clare measures qubit $C$ in basis
$|\pm\rangle_{C}=(|0\rangle_{C}\pm|1\rangle_{C})/\sqrt{2}$ and
finds an outcome ``$+$'', a maximally entangled Bell state
$|\Phi_{+}\rangle_{AB}=(|00\rangle_{AB}+|11\rangle_{AB})/\sqrt{2}$
is established between Alice and Bob, whereas for outcome ``$-$''
the participants share another Bell state
$|\Phi_{-}\rangle_{AB}=(|00\rangle_{AB}-|11\rangle_{AB})/\sqrt{2}$.
If Clare communicates the measurement outcome to Bob and he
applies a phase flip on his qubit $B$ if the outcome was ``$-$'',
Alice and Bob will share a single entangled state
$|\Phi_{+}\rangle_{AB}$, which they can subsequently use, e.g.,
for quantum teleportation \cite{Bennett_93} or dense coding
\cite{Bennett_92}. Localization of a two-qubit entanglement by
a measurement of a qubit from a three-qubit GHZ state has been realized experimentally
with trapped ions \cite{Roos_04}, whereas assisted teleportation has been
implemented with polarization-entangled photons \cite{Zhao_04}.

While for GHZ state (\ref{GHZ}) the amount of localized
entanglement is maximal for all measurement outcomes there are
other three-qubit pure states for which this is not the case
anymore. The localizability of entanglement of these states can be
characterized by an entanglement quantifier called entanglement of
assistance \cite{DiVincenzo_99} defined as an average entropy of
entanglement that can be localized between qubits $A$ and $B$ by a
projective measurement on qubit $C$, which is maximized over all
the measurements. In practice, a better established quantity
called localizable entanglement is commonly used being a
multipartite \cite{Verstraete_04} and mixed-state \cite{Popp_05}
generalization of the entanglement of assistance, to which we will
therefore refer also in what follows.

We have seen that localizability of entanglement is a key property
for performance of several assisted quantum information protocols
and it stays behind introduction of some entanglement quantifiers
ranging from entanglement of assistance
\cite{DiVincenzo_99,Smolin_05} and localizable entanglement
\cite{Verstraete_04,Popp_05} to entanglement of collaboration
\cite{Gour_06}. Besides that, localizability of maximum
entanglement between any two qubits is an essential feature of
cluster states being a backbone of the measurement-based model of
quantum computation \cite{Raussendorf_01,Gross_08}. What is more,
the possibility to localize maximal entanglement between at least
one pair of qubits in a multiqubit pure state is a necessary
condition for the state to be a universal resource for this model
of quantum computation \cite{VandenNest_06}. From the point of
view of utility of a generic multipartite entangled state in
previous applications it is important to know, whether the state
contains localizable entanglement. Although a complete
characterization of the set of states with localizable
entanglement is a daunting task even for three-qubit states, it
might still be possible to draw some conclusions about its
structure. Clearly, the first logical step is to elucidate a more
simple question as to whether a non-empty complement of the set of
three-qubit states with localizable entanglement exists, i.e.,
whether there are some three-qubit entangled states for which
two-qubit entanglement cannot be localized by any measurement on
the remaining qubit.
%Obviously, such states are not a resource for
%all previous protocols as their entanglement is confined such that
%in order to localize it between a pair of qubits a collective
%operation on a pair of qubits is needed. In this respect, states
%with nonlocalizable entanglement share similarity with bound
%entanglement from which maximal pure-state entanglement cannot be
%extracted by any LOCC, but as we will see there are states with
%nonlocalizable entanglement which are not bound entangled.
%As such states possess no separable splitting a different line of
%argumentation has to be used to prove the presence of
%nonlocalizable entanglement.

Apparently, one can find some trivial examples of tripartite
entangled states with nonlocalizable entanglement. For instance,
if a tripartite entangled state is separable across some bipartite
splitting, then its entanglement cannot be localized by any
measurement on a part belonging to the bipartite part of the
splitting. Similarly, if the latter state is separable with
respect to at least two bipartite splittings its entanglement
cannot be localized by any measurement on any of its parts. In all
these cases, nonlocalizability of entanglement occurs in partially
entangled states and it is a direct consequence of the
separability properties of the respective states. Recently,
tripartite entangled states with one and two separable bipartite
splittings have been demonstrated experimentally both for qubits
\cite{Fedrizzi_13} as well as for Gaussian states
\cite{Peuntinger_13,Croal_15,Vollmer_13}.

\begin{figure}[!h]
%%%%%%%%%%%%%%%%%%%%%%%%%%%%%%%%%%%%%%%%%%%%%%%%%%%%%%%%%%%%%%%%%%%%
\centering
\includegraphics[width=0.70\linewidth,angle=00]{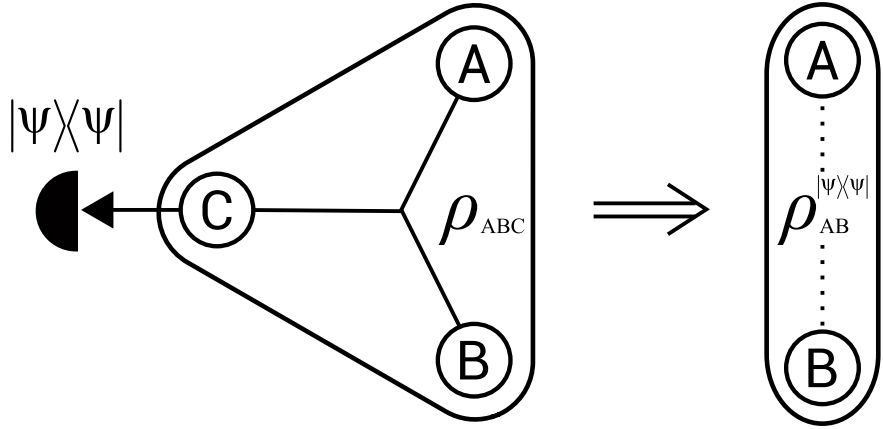}
\caption{A scheme depicting nonlocalizability of entanglement in a fully inseparable state $\rho_{ABC}$ of three qubits $A,B$ and $C$. For any projection of qubit $C$ onto a pure state
$|\psi\rangle$ the resulting two-qubit state $\rho_{AB}^{|\psi\rangle\langle\psi|}$ is separable. See text for details.}
\label{fig0}
\end{figure}

In this paper we investigate both theoretically and experimentally
the little explored area of states with {\it nonlocalizable}
entanglement. Unlike above-mentioned partially entangled states
with nonlocalizable entanglement, here we are interested in
existence of nonlocalizable entanglement in the from the point of
view of applications most important class of fully inseparable
three-qubit states. Out of these states we primarily focus on
states for which entanglement cannot be localized between {\it
any} pair of qubits by any measurement on the third qubit. Instead
of analyzing states with zero localizable entanglement, we analyze
a subset of this set of states given by states for which
entanglement cannot be localized probabilistically by any
measurement (see Fig.~\ref{fig0}). Very recently, an example of
such a state has been constructed \cite{Miklin_16} in the context
of determination of genuine multipartite entanglement only from
two-qubit reductions of the state. The state of ref.~\citen{Miklin_16}
is a genuine three-qubit entangled state with nonlocalizable entanglement,
for which entanglement can be certified only from its two-qubit reductions.
Since the state embodies two properties, it is rather complex and a natural
question arises, whether nonlocalizable entanglement itself can be demonstrated with
less complicated fully inseparable states. Here, we show that
this is really the case. We prove, that nonlocalizable entanglement exists
already in suitable three-qubit convex mixtures of GHZ state
(\ref{GHZ}) and a maximally mixed state. We further generalize the latter set
of states with nonlocalizable entanglement to a two-parametric family of states
by adding a classically correlated fully separable state to the GHZ state.
Next, we show that one can localize both conditionally as well as unconditionally the
nonlocalizable entanglement with the help of the collective
controlled-not (CNOT) operation. Moreover, we also prepare
experimentally a state from the family using single photons and
linear optical circuit and we prove that its entanglement cannot
be localized by any measurement on any of the qubits. Finally, we
conclude our experimental analysis of nonlocalizable entanglement
by extracting a two-qubit entanglement from the experimentally
prepared state using CNOT operation and postselection.

%%%%%%%%%%%%%%%%%%%%%%%%%%%%%%%%%%%%%%%%%%%%%%%%%%%%%%%%%%%%%%%%%%%%%%%%%%%%%%%%%%%%%%%%%%%
%%%%%%%%%%%%%%%%%%%%%%%%%%%%%%%%%%%%%%%%%%%%%%%%%%%%%%%%%%%%%%%%%%%%%%%%%%%%%%%%%%%%%%%%%%%

%The paper is organized as follows. Section~\ref{Sec_2} deals with construction of fully inseparable three-qubit states with nonlocalizable entanglement. In Sec.~\ref{Sec_3} we show how the nonlocalizable entanglement can be concentrated conditionally as well as unconditionally between two qubits using a collective two-qubit CNOT operation. Finally, Sec.~\ref{Sec_4} contains experimental results demonstrating nonlocalizable entanglement and conclusions are drawn in Sec.~\ref{Sec_5}.

\section*{Results}
\subsection*{Fully inseparable three-qubit state with nonlocalizable entanglement}\label{Sec_2}

The presence of nonlocalizable entanglement in a fully inseparable
three-qubit state can be demonstrated on a very simple example of
a convex mixture of the GHZ state (\ref{GHZ}) and the maximally
mixed state $(1/8)\mathds{1}$,
%%%%%%%%%%%%%%%%%%%%%%%%%%%%%%%%%%%%%%%%%%%%%%%%%%%%%%%%%%%%%%%%%%%%%%%%%%%%
\begin{equation}
\label{rhop} \rho_p=p\ket{GHZ}\bra{GHZ}+\frac{1-p}{8}\mathds{1}.
\end{equation}
%%%%%%%%%%%%%%%%%%%%%%%%%%%%%%%%%%%%%%%%%%%%%%%%%%%%%%%%%%%%%%%%%%%%%%%%%%%%
Here, the parameter $p \in [0,1]$ controls the ratio between the
two states, the symbol $\mathds{1}$ denotes the $8\times8$
identity matrix, and we have suppressed the qubit indexes for
brevity. Since the state is symmetric under the exchange of any
two qubits, it is sufficient to investigate the presence of
entanglement and its localizability only with respect to one
bipartite splitting, say $C-(AB)$ splitting. For certification of
entanglement we can use the partial transposition criterion
\cite{Peres_96,Horodecki_96} according to which in order qubit $C$
to be entangled with a pair of qubits $(AB)$ it is sufficient if
the partial transpose of density matrix (\ref{rhop}) with respect
to the qubit ($\equiv\rho_{p}^{T_{C}}$) has a negative eigenvalue.
Because state (\ref{rhop}) belongs to the class of three-qubit
generalizations of the Werner state \cite{Werner_89,Dur_99}, the
condition is also necessary and it reveals that the state is
entangled if and only if $p>1/5$ \cite{Dur_99}.

Moving to the analysis of the localizability of entanglement carried by state (\ref{rhop}) let us assume that the last qubit $C$ is projected onto a pure single-qubit state
%%%%%%%%%%%%%%%%%%%%%%%%%%%%%%%%%%%%%%%%%%%%%%%%%%%%%%%%%%%%%%%%%%%%%%%%%%%
\begin{equation}\label{psi}
|\psi(\vartheta,\phi)\rangle=\cos\left(\frac{\vartheta}{2}\right)|0\rangle+e^{i\phi}\sin\left(\frac{\vartheta}{2}\right)|1\rangle,
\end{equation}
%%%%%%%%%%%%%%%%%%%%%%%%%%%%%%%%%%%%%%%%%%%%%%%%%%%%%%%%%%%%%%%%%%%%%%%%%%%
where $\vartheta\in[0,\pi]$ and $\phi\in[0,2\pi)$. By calculating
the (unnormalized) conditional state of qubits $A$ and $B$,
$\tilde{\rho}_{AB}(\vartheta,\phi)\equiv\mbox{Tr}_{C}[|\psi(\vartheta,\phi)\rangle_{C}\langle\psi(\vartheta,\phi)|\rho_{p}]$,
and normalizing it properly one finds, that the conditional state
reads explicitly as
%%%%%%%%%%%%%%%%%%%%%%%%%%%%%%%%%%%%%%%%%%%%%%%%%%%%%%%%%%%%%%%%%%%%%%%%%%%
\begin{eqnarray}\label{rhoAB}
\rho_{AB}(\vartheta,\phi)&=&p\tau_{AB}(\vartheta,\phi)+\frac{1-p}{4}\tilde{\mathds{1}},
\end{eqnarray}
%%%%%%%%%%%%%%%%%%%%%%%%%%%%%%%%%%%%%%%%%%%%%%%%%%%%%%%%%%%%%%%%%%%%%%%%%%%
where
%%%%%%%%%%%%%%%%%%%%%%%%%%%%%%%%%%%%%%%%%%%%%%%%%%%%%%%%%%%%%%%%%%%%%%%%%%%
\begin{eqnarray}\label{tauAB}
\tau_{AB}(\vartheta,\phi)&=&\cos^{2}\left(\frac{\vartheta}{2}\right)|00\rangle\langle00|+\sin^{2}\left(\frac{\vartheta}{2}\right)|11\rangle\langle11|
+\sin\left(\frac{\vartheta}{2}\right)\cos\left(\frac{\vartheta}{2}\right)\left[e^{i\phi}|00\rangle\langle11|+e^{-i\phi}|11\rangle\langle00|\right]
\end{eqnarray}
%%%%%%%%%%%%%%%%%%%%%%%%%%%%%%%%%%%%%%%%%%%%%%%%%%%%%%%%%%%%%%%%%%%%%%%%%%%
%%%%%%%%%%%%%%%%%%%%%%%%%%%%%%%%%%%%%%%%%%%%%%%%%%%%%%%%%%%%%%%%%%%%%%%%%%%
%\begin{eqnarray}\label{rhoAB}
%\rho_{AB}(\vartheta,\phi)&=&p\left\{\cos^{2}\left(\frac{\vartheta}{2}\right)|00\rangle\langle00|+\sin^{2}\left(\frac{\vartheta}{2}\right)|11\rangle\langle11|\right.\nonumber\\
%&&\left.+\sin\left(\frac{\vartheta}{2}\right)\cos\left(\frac{\vartheta}{2}\right)\left[e^{i\phi}|00\rangle\langle11|\right.\right.\nonumber\\
%&&\left.\left.+e^{-i\phi}|11\rangle\langle00|\right]\right\}+\frac{1-p}{4}\tilde{\mathds{1}},
%\end{eqnarray}
%%%%%%%%%%%%%%%%%%%%%%%%%%%%%%%%%%%%%%%%%%%%%%%%%%%%%%%%%%%%%%%%%%%%%%%%%%%
and $\tilde{\mathds{1}}$ is the $4\times 4$ identity matrix.
Provided that there is $p$ satisfying $1\geq p>1/5$ such that the
conditional state (\ref{rhoAB}) contains no entanglement for any
$\vartheta$ and $\phi$, the state (\ref{rhop}) is a sought example
of a fully inseparable state with nonlocalizable entanglement. To
show that such the $p$ really exists we use the two-qubit
separability criterion which says that a density matrix
$\rho_{AB}$ is separable if and only if
$\mbox{det}(\rho_{AB}^{T_{A}})\geq0$ \cite{Augusiak_08}. Applying
the criterion to state (\ref{rhoAB}) one finds after some algebra
that
%%%%%%%%%%%%%%%%%%%%%%%%%%%%%%%%%%%%%%%%%%%%%%%%%%%%%%%%%%%%%%%%%%%%%%%%%%%
\begin{eqnarray}\label{detrhoABTA}
\mbox{det}\left[\rho_{AB}^{T_A}(\vartheta,\phi)\right]&=&-\frac{p^4}{16}\sin^{4}(\vartheta)-\frac{p^3(1-p)}{16}\sin^{2}(\vartheta)+\left(\frac{1-p}{4}\right)^{3}\left(\frac{3p+1}{4}\right)
\geq\left(\frac{1+p}{4}\right)^{3}\left(\frac{1-3p}{4}\right).
\end{eqnarray}
%%%%%%%%%%%%%%%%%%%%%%%%%%%%%%%%%%%%%%%%%%%%%%%%%%%%%%%%%%%%%%%%%%%%%%%%%%%
Hence we see, that if $p\leq1/3$ the determinant is non-negative
and thus the conditional state (\ref{rhoAB}) is separable for any
$\vartheta$ and $\phi$. On the other hand, the lower bound on the
RHS of inequality (\ref{detrhoABTA}) is
saturated for $\vartheta=\pi/2$ and therefore in this case the
determinant is negative for any $p>1/3$. Thus we have arrived to
the finding that for state (\ref{rhop}) it is impossible to
probabilistically localize entanglement between any pair of qubits
by any projective measurement on the third qubit if and only if
$p\leq1/3$. Since nonlocalizability of entanglement by any
projective measurement implies its nonlocalizability by any
generalized measurement (or even operation) \cite{Miklin_16} we
can conclude, that states (\ref{rhop}), where parameter $p$ lies
in the interval $1/5<p\leq1/3$ are fully inseparable and their
entanglement cannot be localized by any measurement.

A larger set of three-qubit fully inseparable states with nonlocalizable entanglement is obtained if the GHZ state on the RHS of Eq.~(\ref{rhop}) is replaced with the following convex
mixture of GHZ state (\ref{GHZ}) and a classically correlated separable state,
%%%%%%%%%%%%%%%%%%%%%%%%%%%%%%%%%%%%%%%%%%%%%%%%%%%%%%%%%%%%%%%%%%%%%%%%%%%%%%%
\begin{eqnarray}\label{rhomu}
\rho_{\mu}&=&\mu\ket{GHZ}\bra{GHZ}+\frac{1-\mu}{3}(\ket{001}\bra{001}+\ket{010}\bra{010}+\ket{100}\bra{100}),
\end{eqnarray}
%%%%%%%%%%%%%%%%%%%%%%%%%%%%%%%%%%%%%%%%%%%%%%%%%%%%%%%%%%%%%%%%%%%%%%%%%%%
where $\mu \in [0,1]$. This gives the following two-parametric family of mixed three-qubit states
%%%%%%%%%%%%%%%%%%%%%%%%%%%%%%%%%%%%%%%%%%%%%%%%%%%%%%%%%%%%%%%%%%%%%%%%%%%
\begin{equation}\label{rhopmu}
\rho_{p,\mu}=p \rho_\mu+\frac{1-p}{8}\mathds{1}.
\end{equation}
%%%%%%%%%%%%%%%%%%%%%%%%%%%%%%%%%%%%%%%%%%%%%%%%%%%%%%%%%%%%%%%%%%%%%%%%%%%
The state is again invariant under the exchange of any two qubits
and thus also in this case it is sufficient to investigate whether
the state is entangled and whether the entanglement is localizable
only for $C-(AB)$ splitting. Except for the extreme case when
$\mu=1$ state (\ref{rhopmu}) is not a three-qubit generalization
of the Werner state \cite{Dur_99} anymore and the negativity of
the partial transpose $\rho_{p,\mu}^{T_{C}}$ is known to be only
sufficient for the presence of entanglement with respect to the
splitting. Making use of the criterion one finds after some
algebra that partial transposition $\rho_{p,\mu}^{T_{C}}$ of state
(\ref{rhopmu}) possesses seven nonnegative eigenvalues and one
eigenvalue equal to
%%%%%%%%%%%%%%%%%%%%%%%%%%%%%%%%%%%%%%%%%%%%%%%%%%%%%%%%%%%%%%%%%%%%%%%%%%%%%%%%
\begin{eqnarray}\label{alpha}
\alpha=\frac{1}{24}\left[3+p-4p\mu-4p\sqrt{1+2\mu(5\mu-1)}\right]
\end{eqnarray}
%%%%%%%%%%%%%%%%%%%%%%%%%%%%%%%%%%%%%%%%%%%%%%%%%%%%%%%%%%%%%%%%%%%%%%%%%%%%%%%%%
which can be nonnegative or negative. The state (\ref{rhopmu}) contains entanglement across $C-(AB)$ splitting and therefore
it is fully inseparable due to the symmetry, if $\alpha<0$ which is equivalent with the following inequality:
%%%%%%%%%%%%%%%%%%%%%%%%%%%%%%%%%%%%%%%%%%%%%%%%%%%%%%%%%%%%%%%%%%%%%%%%%%%
\begin{equation}\label{pPPT}
p>\frac{3}{4\mu-1+4\sqrt{1+2\mu(5\mu-1)}}\equiv p_{\rm PPT}.
\end{equation}
%%%%%%%%%%%%%%%%%%%%%%%%%%%%%%%%%%%%%%%%%%%%%%%%%%%%%%%%%%%%%%%%%%%%%%%%%%%

\begin{figure}[h]
%%%%%%%%%%%%%%%%%%%%%%%%%%%%%%%%%%%%%%%%%%%%%%%%%%%%%%%%%%%%%%%%%%%%
\centering
\includegraphics[width=0.70\linewidth,angle=00]{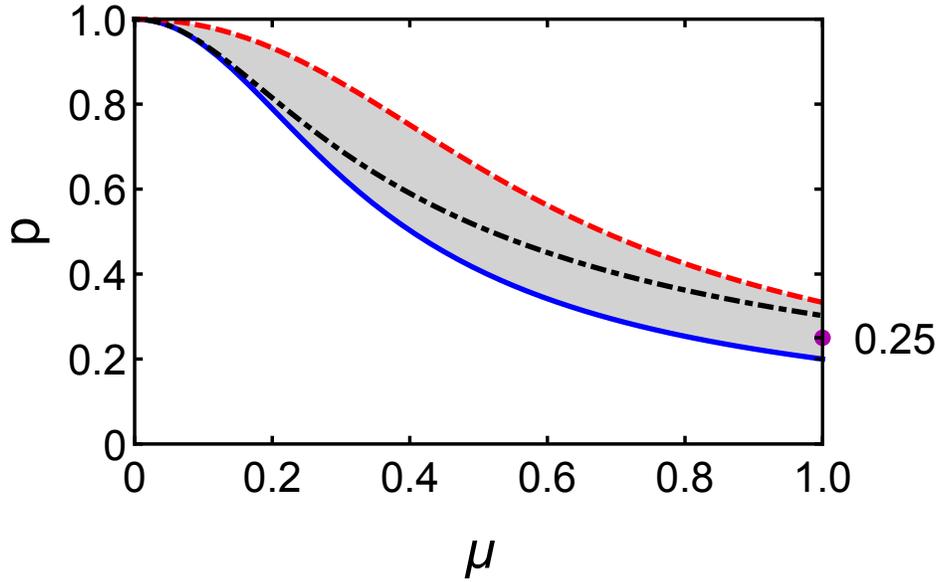}
\caption{Dependence of $p_{\rm PPT}$, Eq.~(\ref{pPPT}), (solid
blue line), $p_{\rm nloc}$, Eq.~(\ref{pnloc}), (dashed red line)
and $p_{\rm col}$, Eq.~(\ref{pcol}), (dashed-dotted black line),
as a function of parameter $\mu$. Gray region depicts fully
inseparable states (\ref{rhopmu}) with nonlocalizable
entanglement. The entanglement of all states above solid blue line
can be transformed into two-qubit entanglement by the CNOT
operation followed by conditioning on outcome ``0'' of the
measurement in computational basis of output target qubit. The
entanglement of all states above dashed-dotted black line can be
transformed into two-qubit entanglement by the CNOT operation
followed by trace-preserving operation (\ref{O}).
A dark magenta point on the right vertical axis depicts a state with nonlocalizable entanglement
$\rho_{1/4,1}$, Eq.~(\ref{rhopmu}), which we prepared experimentally.
See text for details.} \label{fig1}
\end{figure}

Localizability of entanglement carried by states (\ref{rhopmu}) can be investigated analogously as in the case of states (\ref{rhop}). After projection of last qubit $C$ onto vector
(\ref{psi}) density matrix (\ref{rhopmu}) collapses into the state
%%%%%%%%%%%%%%%%%%%%%%%%%%%%%%%%%%%%%%%%%%%%%%%%%%%%%%%%%%%%%%%%%%%%%%%%%%%
\begin{eqnarray}\label{sigmaAB}
\sigma_{AB}(\vartheta,\phi)&=&\frac{1}{p(\vartheta)}\left\{\frac{p\mu}{2}\tau_{AB}(\vartheta,\phi)+\left(\frac{1-p}{8}\right)\tilde{\mathds{1}}
+p\left(\frac{1-\mu}{3}\right)\left[\sin^{2}\left(\frac{\vartheta}{2}\right)|00\rangle\langle00|\right.\right.\nonumber\\
&&\left.\left.+\cos^{2}\left(\frac{\vartheta}{2}\right)\left(|01\rangle\langle01|+|10\rangle\langle10|\right)\right]\right\},
\end{eqnarray}
%%%%%%%%%%%%%%%%%%%%%%%%%%%%%%%%%%%%%%%%%%%%%%%%%%%%%%%%%%%%%%%%%%%%%%%%%%%
where state $\tau_{AB}(\vartheta,\phi)$ is given in Eq.~(\ref{tauAB}) and
%%%%%%%%%%%%%%%%%%%%%%%%%%%%%%%%%%%%%%%%%%%%%%%%%%%%%%%%%%%%%%%%%%%%%%%%%%%%
\begin{equation}\label{ptheta}
p(\vartheta)=\frac{1}{2}\left[1+p\left(\frac{1-\mu}{3}\right)\cos(\vartheta)\right]
\end{equation}
%%%%%%%%%%%%%%%%%%%%%%%%%%%%%%%%%%%%%%%%%%%%%%%%%%%%%%%%%%%%%%%%%%%%%%%%%%%
is a probability of detecting the state (\ref{psi}). In order to identify the region of parameters $p$ and $\mu$
for which state (\ref{sigmaAB}) possesses no entanglement for any $\vartheta$ and $\phi$, we again use the partial
transposition criterion. Instead of investigating the sign of eigenvalues of matrix $\sigma_{AB}^{T_{A}}(\vartheta,\phi)$, it is more convenient to
investigate the sign of eigenvalues of matrix $\tilde{\sigma}_{AB}^{T_{A}}(\vartheta,\phi)=p(\vartheta)\sigma_{AB}^{T_{A}}(\vartheta,\phi)$,
which possess the same sign owing to the inequality $p(\vartheta)>0$. The matrix $\tilde{\sigma}_{AB}^{T_{A}}(\vartheta,\phi)$ has three nonnegative eigenvalues and one eigenvalue
%%%%%%%%%%%%%%%%%%%%%%%%%%%%%%%%%%%%%%%%%%%%%%%%%%%%%%%%%%%%%%%%%%%%%%%%%%%%%%%%
\begin{eqnarray}\label{beta}
\beta=\frac{1}{24}\left\{3+p-4p\mu+2p\left[2(1-\mu)\cos(\vartheta)-3\mu\sin(\vartheta)\right]\right\},\nonumber\\
\end{eqnarray}
%%%%%%%%%%%%%%%%%%%%%%%%%%%%%%%%%%%%%%%%%%%%%%%%%%%%%%%%%%%%%%%%%%%%%%%%%%%%%%%%%
which can be nonnegative or negative. By solving extremal equation $d\beta/d\vartheta=0$ one finds that in the interior of the interval $\vartheta\in[0,\pi]$ eigenvalue (\ref{beta}) has one
stationary point $(\equiv\vartheta_{\rm opt})$ which satisfies equation
%%%%%%%%%%%%%%%%%%%%%%%%%%%%%%%%%%%%%%%%%%%%%%%%%%%%%%%%%%%%%%%%%%%%%%%%%%%%%%%%
\begin{equation}
\label{tg} \tan{(\vartheta_{\rm
opt})}=-\frac{3}{2}\left(\frac{\mu}{1-\mu}\right).
\end{equation}
%%%%%%%%%%%%%%%%%%%%%%%%%%%%%%%%%%%%%%%%%%%%%%%%%%%%%%%%%%%%%%%%%%%%%%%%%%%%%%%%
Hence, we get
%%%%%%%%%%%%%%%%%%%%%%%%%%%%%%%%%%%%%%%%%%%%%%%%%%%%%%%%%%%%%%%%%%%%%%%%%%%%%%%%
\begin{equation}\label{thetaopt}
\vartheta_{\rm
opt}=\pi-\arctan\left[\frac{3}{2}\left(\frac{\mu}{1-\mu}\right)\right]
\end{equation}
%%%%%%%%%%%%%%%%%%%%%%%%%%%%%%%%%%%%%%%%%%%%%%%%%%%%%%%%%%%%%%%%%%%%%%%%%%%%%%%%
which gives
%%%%%%%%%%%%%%%%%%%%%%%%%%%%%%%%%%%%%%%%%%%%%%%%%%%%%%%%%%%%%%%%%%%%%%%%%%%%%%%%
\begin{eqnarray}\label{betaopt}
\beta_{\rm
opt}=\frac{1}{24}\left[3+p-4p\mu-2p\sqrt{4+\mu(13\mu-8)}\right].
\end{eqnarray}
%%%%%%%%%%%%%%%%%%%%%%%%%%%%%%%%%%%%%%%%%%%%%%%%%%%%%%%%%%%%%%%%%%%%%%%%%%%%%%%%%
Comparison of the extremal eigenvalue with values of the eigenvalue (\ref{beta}) at the boundary points $\vartheta=0$ and $\vartheta=\pi$ of the interval $[0,\pi]$ reveals, that it is not
higher than the boundary values, and it is thus a global minimum on the interval. From condition $\beta_{\rm opt}\geq0$ we find, that the conditional state (\ref{sigmaAB})
is separable for any $\vartheta$ and $\phi$ if and only if
%%%%%%%%%%%%%%%%%%%%%%%%%%%%%%%%%%%%%%%%%%%%%%%%%%%%%%%%%%%%%%%%%%%%%%%%%%%
\begin{equation}\label{pnloc}
p\leq\frac{3}{4\mu-1+2\sqrt{4+\mu(13\mu-8)}}\equiv p_{\rm nloc}.
\end{equation}
%%%%%%%%%%%%%%%%%%%%%%%%%%%%%%%%%%%%%%%%%%%%%%%%%%%%%%%%%%%%%%%%%%%%%%%%%%%
Consequently, state (\ref{rhopmu}) for which parameter $p$ satisfies inequalities $p_{\rm PPT}<p\leq p_{\rm nloc}$ is fully inseparable and its
entanglement cannot be localized by any
measurement. The region in the $(\mu,p)$-plane of states (\ref{rhopmu}) with the nonlocalizable entanglement is depicted by a gray color in Fig.~\ref{fig1}.
%%%%%%%%%%%%%%%%%%%%%%%%%%%%%%%%%%%%%%%%%%%%%%%%%%%%%%%%%%%%%%%%%%%%%%%%%%%%%%%%%%%%%%%%%%%
%%%%%%%%%%%%%%%%%%%%%%%%%%%%%%%%%%%%%%%%%%%%%%%%%%%%%%%%%%%%%%%%%%%%%%%%%%%%%%%%%%%%%%%%%%%

\subsection*{Entanglement localization by a collective operation}\label{Sec_3}

As we have already mentioned, the impossibility to localize
entanglement of a tripartite quantum state by any projective
measurement implies, that one cannot do that neither by any
generalized measurement nor even by any probabilistic local
operation \cite{Miklin_16}. This means, that in order to transform
nonlocalizable entanglement in states (\ref{rhopmu}) into
two-qubit entanglement, local action on one qubit and classical
communication to the locations of the other two qubits do not
suffice, and some collective operation on several qubits is
needed. In ref.~\citen{Cubitt_03} it was shown, that a two-qubit
entanglement can be localized from a three-qubit partially
entangled state by the CNOT operation followed by a suitable
measurement on one output qubit or by a suitable trace-preserving
operation on both output qubits. Inspired by this approach we show
in the following subsection, that for all states with
nonlocalizable entanglement investigated in previous section one
can conditionally localize entanglement between qubits $A$ and $B$
by first letting qubit $C$ to interact via the CNOT operation with
qubit $B$, measuring qubit $C$ in computational basis, and
postselecting on projection onto state $|0\rangle$. The next
subsection then deals with unconditional localization of
entanglement between qubits $A$ and $B$, which is reached by
replacing the measurement with a suitable trace-preserving
operation on qubits $B$ and $C$.

\subsubsection*{Conditional localization}\label{Subsec_A}

Let us consider the CNOT operation described by the following
unitary transformation,
%%%%%%%%%%%%%%%%%%%%%%%%%%%%%%%%%%%%%%%%%%%%%%%%%%%%%%%%%%%%%%%%%%%%%%%%
\begin{eqnarray}\label{CNOT}
|00\rangle\rightarrow|00\rangle,\quad
|01\rangle\rightarrow|01\rangle,\quad
|10\rangle\rightarrow|11\rangle,\quad
|11\rangle\rightarrow|10\rangle,
\end{eqnarray}
%%%%%%%%%%%%%%%%%%%%%%%%%%%%%%%%%%%%%%%%%%%%%%%%%%%%%%%%%%%%%%%%%%%%%%%%
where the first and second qubit is called control and target qubit, respectively. Assume, that qubits $B$ and $C$ of state (\ref{rhopmu}) interact via the CNOT operation, where qubit $B$ is
a control qubit and qubit $C$ is a target qubit. The operation then transforms the state to
%%%%%%%%%%%%%%%%%%%%%%%%%%%%%%%%%%%%%%%%%%%%%%%%%%%%%%%%%%%%%%%%%%%%%%%%%%%
\begin{eqnarray}\label{rhoprimed}
\rho_{p,\mu}'=p\mu|\Phi_{+}\rangle\langle\Phi_{+}|\otimes|0\rangle\langle0|+\left(\frac{1-p}{8}\right)\mathds{1}+p\left(\frac{1-\mu}{3}\right)
\left(|001\rangle\langle001|+|011\rangle\langle011|+|100\rangle\langle100|\right),
\end{eqnarray}
%%%%%%%%%%%%%%%%%%%%%%%%%%%%%%%%%%%%%%%%%%%%%%%%%%%%%%%%%%%%%%%%%%%%%%%%%
where $|\Phi_{+}\rangle$ is the maximally entangled Bell state defined below Eq.~(\ref{GHZ}). If we further measure the last qubit $C$ in the computational basis and we find it in state
$|0\rangle$, the obtained (normalized) post-measurement state of qubits $A$ and $B$ attains the form
%%%%%%%%%%%%%%%%%%%%%%%%%%%%%%%%%%%%%%%%%%%%%%%%%%%%%%%%%%%%%%%%%%%%%%%%%%%
\begin{eqnarray}\label{rhocond}
\rho_{AB}^{\rm
c}=\frac{1}{p_{0}}\left[p\mu|\Phi_{+}\rangle\langle\Phi_{+}|+p\left(\frac{1-\mu}{3}\right)|10\rangle\langle10|+\left(\frac{1-p}{8}\right)\tilde{\mathds{1}}\right],
\end{eqnarray}
%%%%%%%%%%%%%%%%%%%%%%%%%%%%%%%%%%%%%%%%%%%%%%%%%%%%%%%%%%%%%%%%%%%%%%%%%%%%
where $p_{0}=[p(4\mu-1)+3]/6$ is the probability of finding state
$|0\rangle$. Like previously we discuss separability properties of
state (\ref{rhocond}) by applying the partial transposition
criterion to unnormalized state $\tilde{\rho}_{AB}^{\rm c}\equiv
p_{0}\rho_{AB}^{\rm c}$. The partial transpose of the latter state
with respect to the first qubit possesses three nonnegative
eigenvalues and one eigenvalue (\ref{alpha}) which is obviously
negative if and only if $p>p_{\rm PPT}$, where $p_{\rm PPT}$ is
defined in inequality (\ref{pPPT}). Hence we see, that for all
entangled states (\ref{rhopmu}) (including states with
nonlocalizable entanglement), i.e., for all states lying above
solid blue line in Fig.~\ref{fig1}, it is indeed possible to
localize entanglement by a CNOT operation on two qubits and
postselection on a suitable outcome of a measurement in
computation basis of one of the output qubits.

\subsubsection*{Unconditional localization}\label{Subsec_B}

There is yet another method of how one can extract nonlocalizable
entanglement of states (\ref{rhopmu}) between two qubits. In
contrast with previous method it is unconditional and it relies on
replacement of the measurement on qubit $C$ behind the CNOT
operation with a suitable trace-preserving completely positive map
on qubits $B$ and $C$. The map is described by the following Kraus
operators \cite{Cubitt_03}:
%%%%%%%%%%%%%%%%%%%%%%%%%%%%%%%%%%%%%%%%%%%%%%%%%%%%%%%%%%%%%%%%%%%%%%%%%%
\begin{eqnarray}\label{O}
\mathcal{O}^{(1)}_{BC}=\mathds{1}_B\otimes|0\rangle_C\langle0|,\quad
\mathcal{O}^{(2)}_{BC}=|0\rangle_B\langle0|\otimes|1\rangle_C\langle1|,\quad
\mathcal{O}^{(3)}_{BC}&=&|0\rangle_B\langle
1|\otimes|1\rangle_C\langle1|,
\end{eqnarray}
%%%%%%%%%%%%%%%%%%%%%%%%%%%%%%%%%%%%%%%%%%%%%%%%%%%%%%%%%%%%%%%%%%%%%%%%%%
which satisfy the trace-preservation condition
$\sum_j\mathcal{O}_{BC}^{(j)\dag}\mathcal{O}_{BC}^{(j)}=\mathds{1}_{BC}$,
where $\mathds{1}_{B}$ ($\mathds{1}_{BC}$) is the single-qubit
(two-qubit) identity matrix on the state space of qubit $B$
(qubits $B$ and $C$). If we transform by the map qubits $B$ and
$C$ of the state after the CNOT operation, Eq.~(\ref{rhoprimed}),
we get
%%%%%%%%%%%%%%%%%%%%%%%%%%%%%%%%%%%%%%%%%%%%%%%%%%%%%%%%%%%%%%%%%%%%%%%%%%
\begin{eqnarray}\label{rhoout}
\rho_{p,\mu}^{\rm
out}&=&\sum_j\mathcal{O}_{BC}^{(j)}\rho_{p,\mu}'\mathcal{O}_{BC}^{(j)\dag}=
p\mu|\Phi_{+}\rangle_{AB}\langle\Phi_{+}|\otimes|0\rangle_{C}\langle0|+p\left(\frac{1-\mu}{3}\right)\left(|100\rangle_{ABC}\langle100|+2|001\rangle_{ABC}\langle001|\right)\nonumber\\
&&+\left(\frac{1-p}{8}\right)\mathds{1}_{A}\otimes\left(\mathds{1}_{B}\otimes|0\rangle_{C}\langle0|+2|01\rangle_{BC}\langle01|\right),
\end{eqnarray}
%%%%%%%%%%%%%%%%%%%%%%%%%%%%%%%%%%%%%%%%%%%%%%%%%%%%%%%%%%%%%%%%%%%%%%%%%%
where $\mathds{1}_{A}$ is the single-qubit identity matrix on the
state space of qubit $A$. Further, by dropping qubit $C$ the
remaining qubits $A$ and $B$ are left in the following state
%%%%%%%%%%%%%%%%%%%%%%%%%%%%%%%%%%%%%%%%%%%%%%%%%%%%%%%%%%%%%%%%%%%%%%%%%%
\begin{eqnarray}\label{rhoout2}
\rho_{AB}^{\rm
out}=p\mu|\Phi_{+}\rangle_{AB}\langle\Phi_{+}|+\left(\frac{1-p}{8}\right)\mathds{1}_{A}\otimes\left(\mathds{1}_{B}+2|0\rangle_{B}\langle0|\right)
+p\left(\frac{1-\mu}{3}\right)\left(|10\rangle_{AB}\langle10|+2|00\rangle_{AB}\langle00|\right).
\end{eqnarray}
%%%%%%%%%%%%%%%%%%%%%%%%%%%%%%%%%%%%%%%%%%%%%%%%%%%%%%%%%%%%%%%%%%%%%%%%%%
There are three nonnegative eigenvalues of the partial transpose
$(\rho_{AB}^{\rm out})^{T_{A}}$ and one eigenvalue
%%%%%%%%%%%%%%%%%%%%%%%%%%%%%%%%%%%%%%%%%%%%%%%%%%%%%%%%%%%%%%%%%%%%%%%%%%%%%%%%
\begin{eqnarray}\label{gamma}
\gamma=\frac{1}{24}\left[6-2p-4p\mu-\sqrt{(3+p-4p\mu)^2+144p^2\mu^2}\right]\nonumber\\
\end{eqnarray}
%%%%%%%%%%%%%%%%%%%%%%%%%%%%%%%%%%%%%%%%%%%%%%%%%%%%%%%%%%%%%%%%%%%%%%%%%%%%%%%%%
%%%%%%%%%%%%%%%%%%%%%%%%%%%%%%%%%%%%%%%%%%%%%%%%%%%%%%%%%%%%%%%%%%%%%%%%%%%%%%%%
%\begin{eqnarray}\label{delta4}
%\delta_4&=&\frac{1}{24}\left\{4p(1-\mu)+6(1-p)\right.\nonumber\\
%&&\left.-\sqrt{[4p(1-\mu)+3(1-p)]^2+144p^2\mu^2}\right\},
%\end{eqnarray}
%%%%%%%%%%%%%%%%%%%%%%%%%%%%%%%%%%%%%%%%%%%%%%%%%%%%%%%%%%%%%%%%%%%%%%%%%%%%%%%%%
which is negative if and only if the parameter $p$ satisfies inequality
%%%%%%%%%%%%%%%%%%%%%%%%%%%%%%%%%%%%%%%%%%%%%%%%%%%%%%%%%%%%%%%%%%%%%%%%%%%
\begin{equation}\label{pcol}
p>\frac{9}{5+4\mu+4\sqrt{1+2\mu(14\mu-1)}}\equiv p_{\rm col}.
\end{equation}
%%%%%%%%%%%%%%%%%%%%%%%%%%%%%%%%%%%%%%%%%%%%%%%%%%%%%%%%%%%%%%%%%%%%%%%%%%%
The lower bound $p_{\rm col}$ is depicted by the black dashed-dotted line in Fig.~\ref{fig1}. The figure unveils that nonlocalizable entanglement of all states lying above the curve
can be unconditionally transformed into two-qubit entanglement via the discussed collective operation on a pair of their qubits.

%%%%%%%%%%%%%%%%%%%%%%%%%%%%%%%%%%%%%%%%%%%%%%%%%%%%%%%%%%%%%%%%%%%%%%%%%%%%%%%%%%%

\subsection*{Experiment}\label{Sec_4}

\begin{figure}[!t!]
%%%%%%%%%%%%%%%%%%%%%%%%%%%%%%%%%%%%%%%%%%%%%%%%%%%%%%%%%%%%%%%%%%%%
\centering
\includegraphics[width=1.00\linewidth,angle=00]{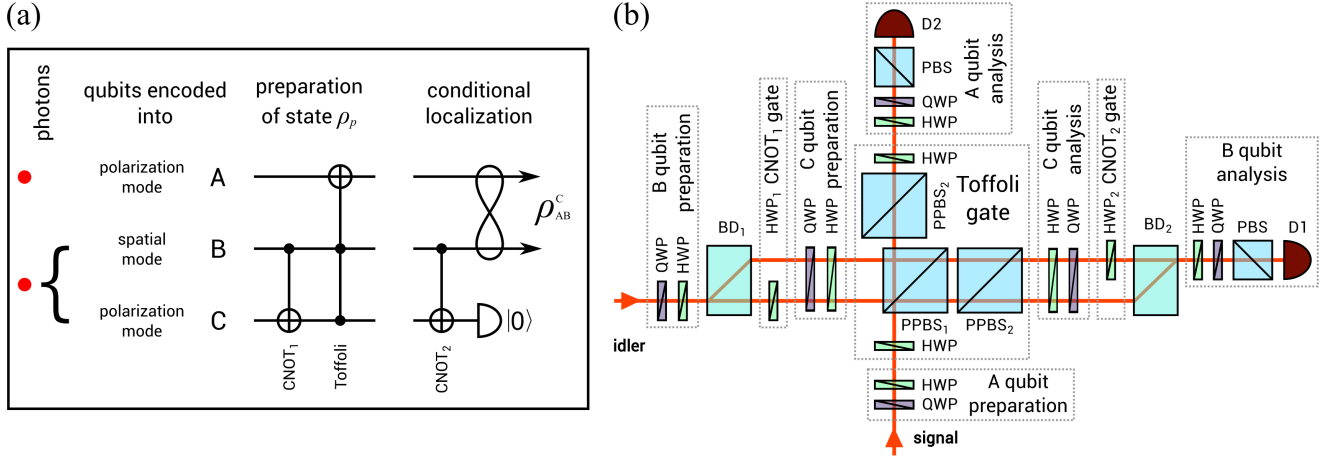}
\caption{Experimental realization of a three-qubit fully
inseparable state with nonlocalizable entanglement. Panel (a)
shows encoding of three qubits $A,B$ and $C$ into two photons by
exploiting their polarization and spatial degrees of freedom, a
circuit for preparation of a state $\rho_{1/4,1}$, Eq.~(\ref{rhopmu}),
with nonlocalizable entanglement consisting of a two-qubit CNOT$_{1}$ gate followed by
a three-qubit Toffoli gate and a circuit for conditional
localization of entanglement between qubits $A$ and $B$ by a
CNOT$_{2}$ gate on qubits $B$ and $C$ followed by a measurement on
qubit $C$. Panel (b) shows a scheme of the experimental setup
implementing state preparation and conditional entanglement
localization from panel (a). The components are labelled as
follows: QWP - quarter-wave plate, HWP - half-wave plate, BD -
calcite beam displacer, PPBS - partially polarizing beam splitter,
PBS - polarizing beam splitter, D - single-photon avalanche diode.
See main text for more details.} \label{fig2}
\end{figure}
%%%%%%%%%%%%%%%%%%%%%%%%%%%%%%%%%%%%%%%%%%%%%%%%%%%%%%%%%%%%%%%%%%%%%%%%%%%%%%%%%%%%%%%%%%%

We have also performed a proof-of-principle experiment demonstrating the existence of a three-qubit fully
inseparable state with nonlocalizable entanglement. We have prepared and analyzed the state $\rho_{1/4,1}$, Eq.~(\ref{rhopmu}), using the
circuits in Fig.~\ref{fig2}~(a) implemented by linear optical setup shown in Fig.~\ref{fig2}~(b). We have used
orthogonally polarized time-correlated photon pairs generated in the process of spontaneous parametric down-conversion in a nonlinear
crystal pumped by a cw laser diode~\cite{Jezek11}. The qubit $A$ is encoded into the polarization degree of freedom of the
signal photon whereas qubits $B$ and $C$  are encoded into the spatial and polarization degree of freedom, respectively, of the idler photon (see Fig.~\ref{fig2}~(a)).

Polarization qubits $A$ and $C$ are prepared and analyzed by
polarization measurement blocks consisting of a quarter-wave plate
(QWP), half-wave plate (HWP), and a polarizing beam splitter (PBS)
where computational states $|0\rangle$ and $|1\rangle$ are
represented by horizontal and vertical polarization, respectively.
The qubit $B$ is initially prepared in polarization encoding using
combination of QWP and HWP and it is subsequently converted into
path encoding using a polarizing beam displacer (BD$_1$). The
computational state $|0\rangle$ corresponds to the horizontally
polarized photon propagating in the upper interferometer arm,
while the state $|1\rangle$ is represented by a vertically
polarized photon propagating in the lower interferometer arm of an
inherently stable Mach-Zehnder interferometer formed by two
calcite beam displacers BD$_1$ and BD$_2$. Polarization to path
conversion produces path-polarization entangled states which can
be disentangled by HWP$_1$ that addresses a single arm of the
interferometer. The action of HWP$_1$ can be regarded as a quantum
CNOT$_{1}$ gate acting on the spatial control and polarization
target qubits. Beam displacer BD$_2$ together with HWP$_2$ map the
spatial qubit back onto polarization one. The core of the setup is
three-qubit Toffoli gate implemented by two-photon interference on
a partially polarizing beam splitter (PPBS$_1$) followed by two
additional PPBSs which serve as partial polarization
filters~\cite{Ralph02,Langford05,Kiesel05,Okamoto05}. Please note,
that Toffoli gate is equivalent to the three-qubit
controlled-controlled-Z gate up to single-qubit Hadamard
transforms on the target qubit (in our case qubit $A$). Our scheme
is probabilistic and operates in coincidence basis where
successful operation is heralded by detection of two-photon
coincidences D1\&D2 at the output. More details about experimental
setup can be found in ref.~\citen{Micuda_15}.

To have full control over the structure of prepared states we have
separately prepared GHZ state (\ref{GHZ}) and all states
$|000\rangle$, $|001\rangle$, ..., $|111\rangle$ of the
computational basis representing diagonal elements of the identity
matrix. To generate GHZ state we have prepared qubit $B$ initially
in $|+\rangle$ state and qubit $C$ in $|0\rangle$ state. With
suitable rotation of HWP$_1$ we have created Bell state
$|\Phi_{+}\rangle_{BC}$ which interacts in Toffoli gate with qubit
$A$ prepared in $|0\rangle$ state. The diagonal elements of the
identity matrix have been prepared in a similar way.

\begin{figure}[!t!]
%%%%%%%%%%%%%%%%%%%%%%%%%%%%%%%%%%%%%%%%%%%%%%%%%%%%%%%%%%%%%%%%%%%%
\centering
\includegraphics[width=1.00\linewidth,angle=00]{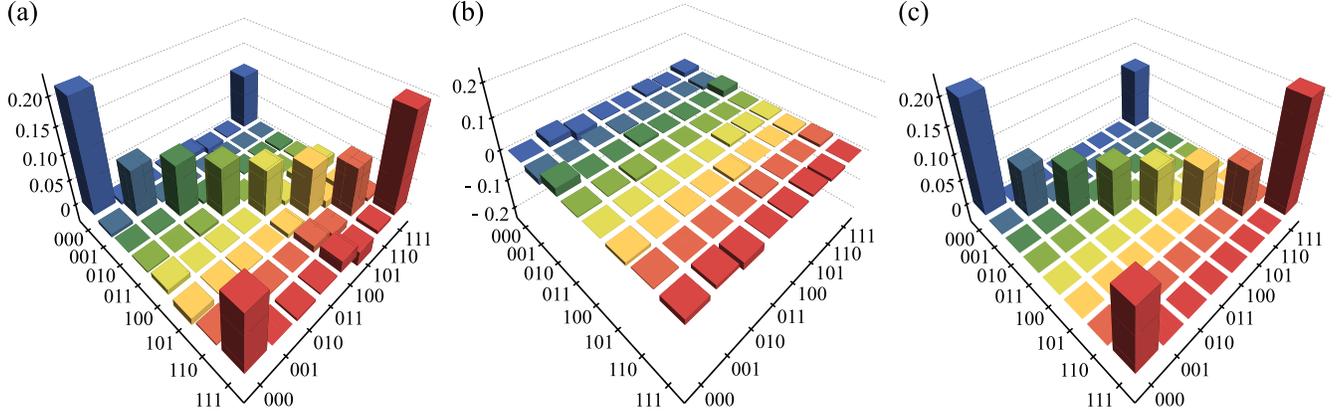}
\caption{Real (a) and imaginary (b) part of the reconstructed density matrix $\rho_{\rm exp}$ and (c) density matrix of the ideal state $\rho_{1/4,1}$. Note that the theoretical density matrix has only real values.} \label{fig3}
\end{figure}

Each prepared state was characterized by three qubit quantum state tomography which consists of sequential projections onto the six states $|0\rangle$, $|1\rangle$, $|+\rangle$, $|-\rangle$, ($|0\rangle + i|1\rangle)/\sqrt{2}$, ($|0\rangle - i|1\rangle)/\sqrt{2}$ at each output qubit for total $6^3=216$ measurements. For each measurement two-photon coincidences were recorded for $50\,\rm{s}$. The measured coincidence counts were normalized by sum of all coincidences and relative frequencies were obtained. The state characterization lasted less than 4 hours. In order to demonstrate nonlocalizability of entanglement on a most simple state, we prepared a three-qubit state~$\rho_{p,\mu}$, Eq.~(\ref{rhopmu}), with $\mu=1$, and to have a sufficiently robust effect against experimental imperfections, we have chosen $p=1/4$, which guarantees that the state lies sufficiently deep inside the set of states with nonlocalizable entanglement (see dark magenta point in Fig.~\ref{fig1}). The relative frequencies of the required state $\rho_{1/4,1}$ have been obtained by mixing relative frequencies of the GHZ state and diagonal elements of the identity matrix, and the state was reconstructed using the maximum likelihood estimation algorithm~\cite{Jezek_03,Hradil_04}. The reconstructed density matrix $(\equiv\rho_{\rm exp})$ exhibits a large overlap with the ideal state $\rho_{1/4,1}$ as is witnessed by a fidelity of
$\mathcal{F}_{\rm exp}(\rho_{\rm exp},\rho_{1/4,1})\equiv\{\mbox{Tr}[(\rho_{\rm exp}^{1/2}\rho_{1/4,1}\rho_{\rm exp}^{1/2})^{1/2}]\}^{2}=0.9841$. For estimation of statistical uncertainty of the experimental results we have used a standard Monte Carlo analysis. Using measured coincidence counts as a mean value of Poisson distribution we have numerically generated 1000 samples of the state $\rho_{\rm exp}$, which were again reconstructed with the help of the maximum likelihood estimation algorithm. For the generated set of density matrices we obtained the average fidelity of $\mathcal{F}_{\rm est}=0.9840\pm0.0002$, which is in an excellent agreement with the experimental value. To get a deeper insight into the structure of the experimentally prepared
state $\rho_{\rm exp}$ and to see its resemblance to the ideal state $\rho_{1/4,1}$ we display both states in Fig.~\ref{fig3}.

%%%%%%%%%%%%%%%%%%%%%%%%%%%%%%%%%%%%%%%%%%%%%%%%%%%%%%%%%%%%%%%%%%%%%%%%%%%%%%%%%%%%%%%%%%%

In the next step of our analysis we certified full inseparability of the prepared state by calculating lowest eigenvalues
$\alpha^{(j)}\equiv\mbox{min}[\mbox{eig}(\rho_{\rm exp}^{T_{j}})]$, $j=A,B,C$, for all three partial transpositions of the
experimental density matrix $\rho_{\rm exp}$. The errors have been again calculated with the help of the Monte Carlo method.
Within the statistical uncertainty the experimental eigenvalues coincide with average eigenvalues
obtained from the computationally generated population of the experimental density matrix. The eigenvalues together
with the errors are summarized in Tab.~\ref{tableI}.
%%%%%%%%%%%%%%%%%%%%%%%%%%%%%%%%%%%%%%%%%%%%%%%%%%%%%%%%%%%%%%%%%%%%%%%%%%%%%%%%%%%%%%%%%%%%%%%%%%%%%%
%\begin{widetext}
\begin{table}[ht]
\caption{Minimal eigenvalue $\alpha^{(j)}$ with one standard deviation of the
partial transpose with respect to qubit $j$ of the experimental density matrix
$\rho_{\rm exp}$. Errors have been obtained using the Monte Carlo method:} \centering
\begin{tabular}{| c | c | c | c |}
\hline $j$ & A & B & C \\
\hline $\alpha^{(j)}\cdot10^2$ & $-3.37\pm0.05$ & $-3.00\pm0.05$ & $-1.98\pm0.05$ \\
\hline
\end{tabular}
\label{tableI}
\end{table}
%\end{widetext}
%%%%%%%%%%%%%%%%%%%%%%%%%%%%%%%%%%%%%%%%%%%%%%%%%%%%%%%%%%%%%%%%%%

\begin{figure}[!t!]
%%%%%%%%%%%%%%%%%%%%%%%%%%%%%%%%%%%%%%%%%%%%%%%%%%%%%%%%%%%%%%%%%%%%
\centering
\includegraphics[width=1.00\linewidth,angle=00]{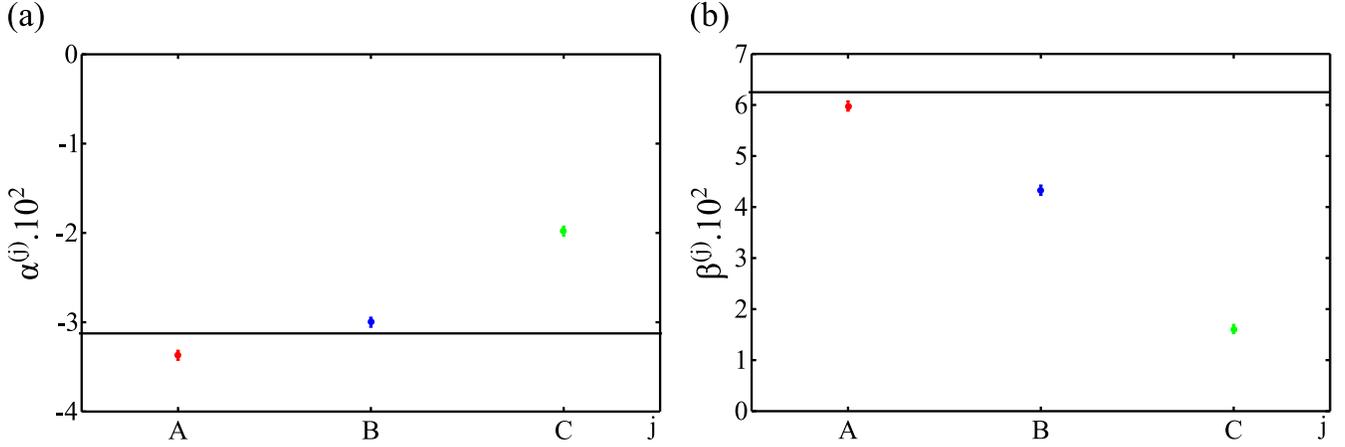}
\caption{Figure (a) displays eigenvalues $\alpha^{(A)}$ (red), $\alpha^{(B)}$ (blue) and $\alpha^{(C)}$ (green) from Tab.~\ref{tableI}.
Solid line in figure (a) depicts the theoretical value of the eigenvalues, $\alpha\doteq-0.031$. Figure (b) displays eigenvalues $\beta^{(A)}$ (red),
$\beta^{(B)}$ (blue) and $\beta^{(C)}$ (green) from Tab.~\ref{tableII}. Solid line in figure (b) represents
the theoretical value of the eigenvalues, $\beta\doteq0.063$. Error bars have been calculated by the Monte Carlo method.
See text for details.} \label{fig4}
\end{figure}

For a better illustration, we display the eigenvalues from Tab.~\ref{tableI} in Fig.~\ref{fig4}~(a).
Inspection of the figure unambiguously proves, that all the eigenvalues are many standard deviations below
zero and therefore the prepared state is fully inseparable by the partial transposition criterion. Note further,
that while according to the theory the eigenvalues should all be equal to a single value
$\alpha=-1/32\doteq-0.031$, Eq.~(\ref{alpha}) with $\mu=1$ and $p=1/4$, they increase as we go from
qubit $A$ to qubit $C$ (see Fig.~\ref{fig4}~(a)). This behaviour can be attributed to the fact that each
qubit passes through different number of optical elements and thus suffers by different state-dependent losses.
Indeed, a simple theoretical model of our operation consisting of an ideal Toffoli gate followed by local filters on
each qubit qualitatively captures the behaviour of experimental eigenvalues and thus confirms
this intuitive explanation.
%The asymmetry of the experimental eigenvalues
%%comes from the fact, that the operation realized by our experiment setup deviates from the ideal Toffoli gate
%as can be seen by applying the tomographically reconstructed matrix of the operation \cite{Micuda_15}
%to ideal input states, which reproduces the experimental eigenvalues in Tab.~\ref{tableI}.

%%%%%%%%%%%%%%%%%%%%%%%%%%%%%%%%%%%%%%%%%%%%%%%%%%%%%%%%%%%%%%%%%%%%%%%%%%%%%%%%%%%%%%%%%%%

In the final step, we have verified nonlocalizability of entanglement in the prepared state again using partial
transposition criterion. First, we have calculated from the experimental density matrix $\rho_{\rm exp}$
the (normalized) conditional state ($\equiv\rho_{{\rm exp}, kl}(\vartheta,\phi)$) of qubits $k$ and $l$,
$k\ne l=A,B,C$, after projection of qubit $j\ne k,l$ of the state $\rho_{\rm exp}$ onto pure state (\ref{psi}).
Next, we have calculate the optimized eigenvalue $\beta^{(j)}\equiv\mathop{\mbox{min}}_{\vartheta,\phi}\{\mbox{min}[\mbox{eig}(\rho_{{\rm exp}, kl}(\vartheta,\phi)^{T_{k}})]\}$,
$j=A,B,C$, by numerical minimization of the lowest eigenvalue of the partial transpose of the conditional state
$\rho_{{\rm exp}, kl}(\vartheta,\phi)$ with respect to qubit $k$ over parameters $\vartheta$ and $\phi$. Further, making use once
again the Monte Carlo analysis we have also calculated averages of the eigenvalues $\beta^{(j)}$ and the
corresponding errors. Like in the previous case, the experimental eigenvalues were found to be equal to
the average eigenvalues within the presented accuracy. The experimental eigenvalues with errors are summarized
in Tab.~\ref{tableII}.

%%%%%%%%%%%%%%%%%%%%%%%%%%%%%%%%%%%%%%%%%%%%%%%%%%%%%%%%%%%%%%%%%%%%%%%%%%%%%%%%%%%%%%%%%%%%%%%%%%%%%%
%\begin{widetext}
\begin{table}[ht]
\caption{Optimized eigenvalue $\beta^{(j)}$ with one standard deviation, which is given by the minimum
over all $\vartheta\in[0,\pi]$ and $\phi\in[0,2\pi)$ of the lowest eigenvalue of the
partial transpose with respect to first qubit of a conditional state obtained by projection of
qubit $j$ of the experimental density matrix $\rho_{\rm exp}$ onto pure state (\ref{psi}).
Errors have been calculated using the Monte Carlo method:}
\centering
\begin{tabular}{| c | c | c | c |}
\hline $j$ & A & B & C \\
\hline $\beta^{(j)}\cdot10^2$ & $5.98\pm0.09$ & $4.33\pm0.09$ &  $1.61\pm0.08$ \\
\hline
\end{tabular}
\label{tableII}
\end{table}
%\end{widetext}
%%%%%%%%%%%%%%%%%%%%%%%%%%%%%%%%%%%%%%%%%%%%%%%%%%%%%%%%%%%%%%%%%%

The eigenvalues from Tab.~\ref{tableII} are displayed in Fig.~\ref{fig4}~(b). The figure reveals
that all the eigenvalues lie by many standard deviations above zero and thus the entanglement
carried by the experimentally prepared state is indeed nonlocalizable by any measurement.
Similar to eigenvalues certifying full inseparability, also the experimental eigenvalues from
Fig.~\ref{fig4}~(b) differ from the theoretical eigenvalue $\beta=1/16\doteq0.063$.
This behaviour can be again attributed to the previously discussed imperfect realization
of Toffoli gate.

We have accomplished experimental investigation of the concept of
nonlocalizable entanglement by extracting conditionally two-qubit
entanglement from the prepared state $\rho_{1/4,1}$ via CNOT$_{2}$
operation and postselection as described in subsection dedicated
to localization of entanglement by a collective operation. The
preparation of state $\rho_{1/4,1}$ was the same as described above.
Next, we have performed CNOT$_{2}$ operation using HWP$_2$ on
qubits $B$ and $C$, where qubit $B$ is a control qubit and qubit
$C$ a target qubit. Finally, we have performed projection of qubit
$C$ onto the $|0\rangle$ basis state. Acquired data were processed
as in the case of preparation of state $\rho_{1/4,1}$. The obtained
conditional two-qubit state $\rho_{AB}^{\rm c}$, Eq.~(\ref{rhocond})
with $p=1/4$ and $\mu=1$, has been again characterized by the
quantum state tomography. The lowest eigenvalue of the partial
transpose $(\rho_{AB}^{\rm c})^{T_{A}}$ of the reconstructed density
matrix $\rho_{AB}^{\rm c}$ reads
$\delta\equiv\mbox{min}\{\mbox{eig}[(\rho_{AB}^{\rm c})^{T_{A}}]\}=-0.036\pm0.001$.
The negativity of the eigenvalue clearly confirms successful
conditional localization of two-qubit entanglement by a collective
operation on a pair of qubits of the prepared state $\rho_{1/4,1}$.
Note finally, that owing to experimental imperfections
the experimental eigenvalue $\delta$ is larger than the ideal theoretical value of the
eigenvalue of $\alpha/p_{0}=-1/20=-0.05$, where we used Eq.~(\ref{alpha}) and the formula for
success probability $p_{0}$ given below Eq.~(\ref{rhocond}) with $\mu=1$ and $p=1/4$.
%Note, that according to theory the latter eigenvalue should
%coincide with the lowest eigenvalue of the
%partial transpose $\rho_{1/4,1}^{T_{A}}$ and they are indeed very close
%to each other as can be seen by comparing the eigenvalue $\delta$ with
%the eigenvalue $\alpha^{(A)}$ from the first column of Tab.~\ref{tableI}.
%A small discrepancy of the two eigenvalues can be attributed to the fact that they have
%been obtained from two different sets of measurement data.

\section*{Discussion}\label{Sec_5}

We have proposed and experimentally demonstrated the concept of nonlocalizable entanglement in the context of three-qubit fully inseparable states.
This type of entanglement is confined in a three-qubit system such, that no measurement on either of the qubits is capable to localize entanglement between the remaining two qubits. Because this property guarantees that also no probabilistic operation can accomplish this \cite{Miklin_16}, there is no way of how one could turn three-qubit entanglement of the state into a useful two-qubit entanglement by using local operation on one qubit and postselecting the other two qubits on successful realization of the operation. In this respect, nonlocalizable entanglement resembles nondistillable (bound) entanglement carried by tripartite states for which three parties holding parts of such a state cannot establish entanglement between any two of them with the help of the third one by LOCC \cite{Dur_99,Dur_00}. In fact, for qubit states localizability of entanglement suffices for distillability. This is because by measuring suitably one qubit of a three-qubit state with localizable entanglement we create entanglement between the remaining two qubits which is always distillable \cite{MHorodecki_97}. Our results further show, that localizability of entanglement is not necessary for distillability as there exist states with nonlocalizable entanglement which are distillable. As an example of such a state can serve us mixtures $\rho_{p}$, Eq.~(\ref{rhop}),
of the GHZ state and a maximally mixed state with $1/5<p\leq1/3$, from which we have prepared the state wit $p={1/4}$ experimentally. According to our results, such states carry nonlocalizable entanglement because they do not surpass the nonlocalizability bound $p_{\rm nloc}=1/3$ and at the same time they are distillable, as all the states with $p>1/5$ are distillable \cite{Dur_99}. The existence of such states is expectable, because we consider localization of entanglement by local operations on one part of a single copy of the state followed by a postselection of the other two parts, whereas for entanglement distillation more powerful LOCC operations acting on all three parts of generally multiple copies of the state are used. The question of the existence of states with single-copy nonlocalizable entanglement which would be localizable when two or more copies are measured jointly
is deferred for further research. Their existence is likely because one can find three-qubit states which give strictly more than two times larger average entropy of entanglement that can be localized by a measurement on two copies of the state than that of one can localize on a single copy \cite{DiVincenzo_99}.

We have seen that extraction of two-qubit entanglement from three-qubit states with nonlocalizable entanglement would require more copies of the state and more powerful LOCC operations. Another option is to work only with a single copy of the state but resign on the LOCC character of the used method. Here, we have proposed for the investigated family of states both conditional and unconditional localization method based on application of the CNOT operation on a pair of qubits of the state followed by a measurement on one output of the operation and trace-preserving operation on both outputs, respectively. Additionally, for the prepared mixture we have also realized the more simple conditional method experimentally. Needles to say finally, that the investigated mixtures $\rho_{p}$ with nonlocalizable entanglement do not possess the strongest form of multipartite entanglement which usually appears in applications. The true is that the states are entangled with respect to all three bipartite splittings and thus they are fully inseparable. On the other hand, the states satisfy a necessary and
sufficient condition for biseparability $p\leq 3/7$ \cite{Guhne_10}, and therefore they can be created by mixing of three-qubit entangled states which are separable across different bipartite splittings. The strongest form of multipartite entanglement, the so called genuine multipartite entanglement, is
carried by states which are not biseparable. To demonstrate the existence of nonlocalizable genuine multipartite entanglement we
would have to leave the set of states $\rho_{p}$ in which the two properties never coexist. To the best of our knowledge, there is
currently known one example of a state which carries simultaneously both nonlocalizable and genuine multipartite
entanglement \cite{Miklin_16}. However, an experimental demonstration of such a state would be much more challenging in
comparison with the states investigated here owing to the need to prepare large coherent superpositions of three-qubit computational
basis states with precisely adjusted absolute values and phases of nontrivial complex amplitudes. We hope that our findings
will stimulate further investigation of multipartite fully
inseparable states which are too noisy to possess localizable
entanglement but which are not noisy enough to be nondistillable.

\section*{Acknowledgements}
This work was supported by the Czech Science Foundation
(GA16-17314S). Martina Mikov\'{a} and Ivo Straka acknowledge
support by Palack\'{y} University (IGA-PrF-2016-009).

\section*{Author contributions statement}
M.M. designed the experimental setup and performed the experiment with contributions from M.J., I.S., and M. Mikov\'{a}.
L.M. performed the theoretical calculations with contributions from D.K. M.M., D.K. and L.M. analyzed the experimental data. L.M. and M.M. wrote the manuscript with input from all authors.

\section*{Additional information}
The authors declare no competing financial interests. Correspondence and requests for material should be addressed to L.M. mista@optics.upol.cz.

\end{document}